# Designing devices for wave-vector manipulation using a transformation-optics approach

**Mircea Giloan**

Company for Applied Informatics, Republicii nr. 107, 400489 Cluj-Napoca, Romania

A transformation optics approach was used to derive a general method for designing electromagnetic devices able to manipulate the wave vectors in the specific manner required by the functionality of the device. While the wave paths inside the device remain of a secondary importance the wave vectors are gradually changed in the desired way by choosing the appropriate coordinate transformation. The proposed method was applied to design both converging and diverging flat lenses. Computer simulations revealed the focusing ability of a converging flat lens designed using the proposed technique.

## I. Introduction

The concept of transformation optics has enabled the design of new devices capable to manipulate the electromagnetic waves in an unprecedented manner. This new approach is based on the theoretical result which shows that Maxwell's equations have the same form whether we apply a coordinate transformation or introduce specific gradients in the permittivity and permeability of the original space [1, 2]. This approach was successfully used to design a variety of electromagnetic devices including electromagnetic cloaks, field concentrators, and lenses [3-8]. Since these devices are performing exotic manipulations of the electromagnetic field, non-trivial transformation functions are applied to the initial coordinates system leading to high anisotropic and inhomogeneous materials in regards of their electromagnetic properties. These requirements can be fulfilled by exploiting the impressive achievements from the field of metamaterials which allow the design of inhomogeneous composites with a precise control over their constitutive parameters [9, 10].

An electromagnetic device alters the wave path and/or the wave vector of an incident wave according to the desired functionality. These changes may occur at the interface of two homogeneous media or inside an anisotropic and inhomogeneous medium. So far, the concept of transformation optics was used to design different devices, mainly by focusing on the manipulation of the wave path of the electromagnetic waves [11, 12]. Since transformation optics approach provides anisotropic and inhomogeneous media, capable to manipulate not only the wave paths but also the wave vectors, useful devices can be designed by targeting the manipulation of the wave vectors inside the medium.

In this study we present the design of a flat device consisting of a medium delimited by two parallel planes. The required functionality of the device is a proper manipulation of the wave vector inside the medium in order to provide a specific wave vector distribution at the output plane of the device. The wave vector distribution at the output plane will determine the propagation path of the wave into the homogeneous medium which surrounds the device. This theoretical study shows a general method to retrieve the proper coordinate transformation, and consequently the corresponding permittivity and permeability of the device, which will generate the desired propagation effect based on the wave vector distribution at the output plane of the device.

These theoretical results are applied for designing flat lenses capable to engineer the wave vector distributions in order to achieve a perfect convergence of the emergent rays into a focal point, while keeping the ray trajectories inside the lens as simple straight lines. The implementation of the designed lenses only requires positive optical parameters when their span is limited to the area of the transformation medium with positive constitutive parameters. The study of reflection and transmission shows that for the designed lenses the reflections play a minor role when the focal point is reasonably far from the output plane of the lens. The designs of the perfect lens and the converging lens proposed in this paper may be considered as complementary from the following reasons: the perfect lens achieves perfect imaging in the near field by manipulating only the ray trajectories and using only negative optical parameters [13], while the proposed converging lens focuses very well in the far field by manipulating only the wave vectors and using only positive optical parameters.

## II. Wave vector manipulation

### A. Transformation optics approach

The schematic representation depicted in figure 1 shows the main geometrical characteristics of the designed device, the trajectories of the wave paths inside the device and after leaving the device, and the components of the wave vectors of the incident and emergent waves. The device consists in a medium delimited by two planes perpendicular on *z*-axis. The wave path of a normally incident wave is constrained to a parallel line to *z*-axis. In order to manipulate the wave vector inside the device the following coordinate transformation of the original space was used

$$\begin{cases} x' = x \\ y' = y \\ z' = z \cdot f(x,y) \end{cases} \quad (1)$$

where $f(x,y)$ is a function of two variables which transforms the z coordinate of the original space with respect to *x* and *y* coordinates. For simplicity of the following equations it is useful to denote the multiplicative inverse of function *f* by *h*, hence $h(x,y) = 1/f(x,y)$. Expressed in terms of the *x'*, *y'*, and *z'* coordinates of the transformed space the Jacobian matrix of this coordinate change has the following form

$$J = \begin{pmatrix} 1 & 0 & 0 \\ 0 & 1 & 0 \\ -z' \dfrac{h_x(x',y')}{h(x',y')} & -z' \dfrac{h_y(x',y')}{h(x',y')} & \dfrac{1}{h(x',y')} \end{pmatrix} \quad (2)$$

where $h_x$ and $h_y$ denote the partial derivatives of the *h* function with respect to *x* and *y* variables, respectively. Although, all the following equations are expressed in terms of the *x'*, *y'*, and *z'* coordinates of the transformed space, for simplicity the prime symbol denoting the transformed coordinates will be dropped. Considering that the original space is the free space the relative electric permittivity (*ε*) and magnetic permeability (*μ*) of the anisotropic medium, having an identical electromagnetic behavior as the transformed space, are given by the same symmetric tensor [14-16]

$$n=\begin{pmatrix} h(x,y) & 0 & -z\,h_x(x,y) \\ 0 & h(x,y) & -z\,h_y(x,y) \\ -z\,h_x(x,y) & -z\,h_y(x,y) & \dfrac{g^2(x,y,z)+1}{h(x,y)} \end{pmatrix} \quad (3)$$

where $g(x,y,z)$ is a function which satisfies the following identity

$$g^2(x,y,z)=z^2(h_x^2(x,y)+h_y^2(x,y)) \quad (4)$$

The parametric description of the path and the wave vector of an electromagnetic wave passing through an inhomogeneous and anisotropic medium can be obtained using a Hamiltonian formalism [17]. The Hamiltonian function derived from the plane wave dispersion relation has the form

$$H = \mathbf{k}^T \cdot n \cdot \mathbf{k} - det(n) \quad (5)$$

where $\mathbf{k}$ denotes the dimensionless wave vector ($\mathbf{k}^T=(k_x,k_y,k_z)$), the superscript $T$ denotes the transpose operator and $det(n)=h(x,y)$ is the determinant of tensor $n$ [18, 19]. As follow, the wave vector refers to the dimensionless wave vector which is the wave vector divided by the factor $\omega/c$ where $\omega$ is the angular frequency of the wave and $c$ is the speed of light in the free space. After computing the multiplications $\mathbf{k}^T \cdot n \cdot \mathbf{k}$ the Hamiltonian function becomes

$$H=(k_x^2+k_y^2-1)h-2zk_z(k_x h_x+k_y h_y)+k_z^2\dfrac{g^2+1}{h} \quad (6)$$

The differential equations describing the evolution of the wave path and the wave vector are given by the Hamilton's canonical equations

$$\dfrac{d\mathbf{r}}{d\tau}=\dfrac{\partial H}{\partial \mathbf{k}} \quad (7.1)$$

$$\dfrac{d\mathbf{k}}{d\tau}=-\dfrac{\partial H}{\partial \mathbf{r}} \quad (7.2)$$

where $\mathbf{r}$ is a position vector describing the trajectory of the wave path ($\mathbf{r}^T=(x,y,z)$) and $\tau$ parametrizes the wave path and the wave vector. Applying the first canonical equation to the Hamiltonian function given by Eq. 6 we obtain the following differential equations which are governing the trajectory of the wave path

$$\dfrac{dx}{d\tau}=2(k_x h - z k_z h_x) \quad (8.1)$$

$$\dfrac{dy}{d\tau}=2(k_y h - z k_z h_y) \quad (8.2)$$

$$\dfrac{dz}{d\tau}=2(k_z\dfrac{g^2+1}{h}-z(k_x h_x+k_y h_y)) \quad (8.3)$$

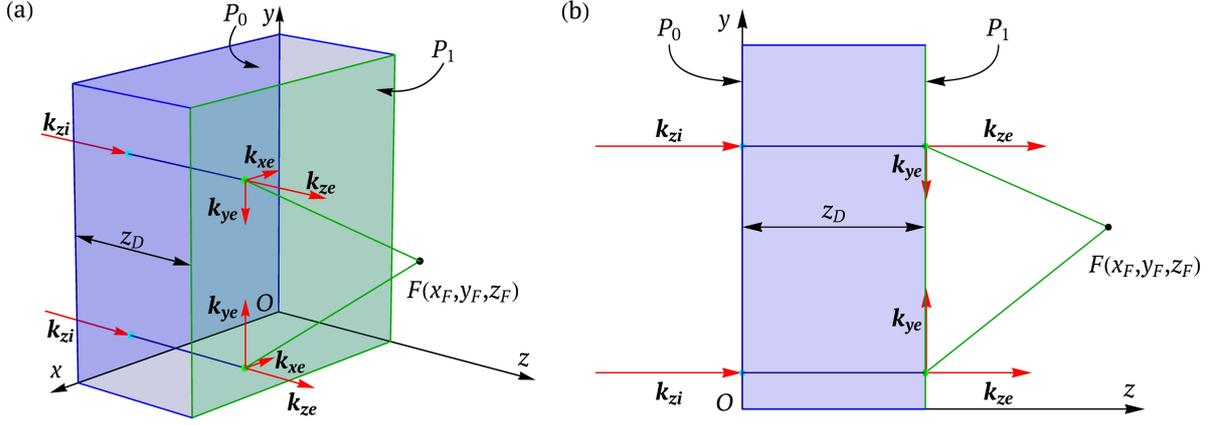

**Figure 1. (a)** 3D view and **(b)** yz plane view, of a reduce area of the device. The device is delimited by the input ($P_0$) and output ($P_1$) planes perpendicular on z-axis and displaced by $z_D$. The wave paths are depicted inside and outside the device by blue and green lines respectively. $F$ denotes the focal point of the device. The incident and emergent wave vector components are depicted by red arrows.

During the propagation of the electromagnetic wave inside the device $x$ and $y$ components of vector $\mathbf{r}$ will remain unchanged and consequently their derivatives with respect to $\tau$ parameter are equal to zero. Hence, the following identities occur from Eqs. 8.1 and 8.2

$$k_x = z\, k_z \frac{h_x}{h} \tag{9.1}$$

$$k_y = z\, k_z \frac{h_y}{h} \tag{9.2}$$

The fulfillment of these identities at the input plane of the device, where $z$ is equal to zero requires for the $k_x$ and $k_y$ components of the wave vector to be also equal to zero at the input plane. Since the law of refraction states that the transverse components of a wave vector conserve across the boundary, the fulfillment of Eqs. 9 at the input plane of the device can be assured only at normal incidence. Hence, the wave vector of the incident wave has the form $\mathbf{k}_i^T = (0,\ 0,\ k_{zi})$. (see Fig. 1). Plugging the identities given by Eq. 4 and Eqs. 9 into the canonical equation of $z$ (Eq. 8.3), the derivative of the $z$ component of the vector $\mathbf{r}$ with respect to the parameter $\tau$ reduces to

$$\frac{dz}{d\tau} = 2\frac{k_z}{h} \tag{10}$$

Considering the identities given by Eq. 4 and Eqs. 9 the Hamiltonian function described by Eq. 6 reduces to the following form

$$H = \frac{k_z^2}{h} - h \tag{11}$$

Using this simplified form of the Hamiltonian function and taking into account that $h$ function is a function of $x$ and $y$ variables having its derivative with respect to $z$ variable equal to zero, the canonical equations of the wave vector components become

$$\frac{dk_x}{d\tau} = \left(\frac{k_z^2}{h^2} + 1\right) h_x \qquad (12.1)$$

$$\frac{dk_y}{d\tau} = \left(\frac{k_z^2}{h^2} + 1\right) h_y \qquad (12.2)$$

$$\frac{dk_z}{d\tau} = 0 \qquad (12.3)$$

Equation 12.3 shows that the *z*-axis component of the wave vector will remain unchanged during the propagation of the electromagnetic wave inside the device. The wave vector at the input plane of the device can be obtain by requiring the fulfillment of the plane wave dispersion relation represented by the Hamiltonian function $H(\mathbf{k}) = 0$ [19]. For the Hamiltonian function given by Eq. 11 we obtain that $k_z = h$. This result simplifies Eqs. 10, 12.2, and 12.3 to

$$\frac{dz}{d\tau} = 2 \qquad (13.1)$$

$$\frac{dk_x}{d\tau} = 2 h_x \qquad (13.2)$$

$$\frac{dk_y}{d\tau} = 2 h_y \qquad (13.3)$$

Dividing Eqs. 13.2 and 13.3 by Eq. 13.1 we obtain the differential equations which are describing the evolution of the *x*- and *y*-axis components of the wave vector with respect to *z* coordinate

$$\frac{dk_x}{dz} = h_x \qquad (14.1)$$

$$\frac{dk_y}{dz} = h_y \qquad (14.2)$$

These equations show that the $k_x$ and $k_y$ components of the wave vector show a linear dependence with respect to *z* coordinate as the wave propagates inside the device. Note that the same linear dependence of $k_x$ and $k_y$ with respect to *z* results from Eqs. 9 when the value of $k_z$ is replaced with *h*. Taking into account that the refraction phenomenon occurring at the output interface of the device conserve the transverse components of the wave vector, the *x*- and *y*-axis components of the wave vector of the emergent wave are

$$k_{xe} = z_D h_x \qquad (15.1)$$
$$k_{ye} = z_D h_y \qquad (15.2)$$

where $z_D$ is the thickness of the device. The *z*-axis component of the emergent wave vector can be obtained by solving the equation $H_0(\mathbf{k}_e) = 0$ where $H_0$ is the Hamiltonian function of the free space $H_0 = \mathbf{k}^T \cdot \mathbf{k} - 1$ and $\mathbf{k}_e$ is the wave vector of the emergent wave [17]. Hence, the components of the emergent wave vector are satisfying the following relation

$$k_{xe}^2 + k_{ye}^2 + k_{ze}^2 = 1 \qquad (16)$$

Eqs. 15 and 16 allow us to retrieve the transformation function $f(x,y)=1/h(x,y)$ of the initial $z$ coordinate which will produce at the output interface of the device the wave vector distribution required by the functionality of the device. Once this function is found the material parameters of the device are easily retrieved using Eq. 3.

### B. Flat lens design

These theoretical results provide the necessary tools for designing flat media able to focalize an incident wave into an arbitrary area. For instance if we want to focus an incident wave into a point $F(x_F, y_F, z_F)$ we need all the wave paths emerging from the medium to pass through this point (see Fig. 1). This requirement is satisfied when the wave vector distribution of the emergent wave fulfills the following equations

$$\frac{k_{xe}}{k_{ze}}=\frac{x-x_F}{z_D-z_F} \tag{17.1}$$

$$\frac{k_{ye}}{k_{ze}}=\frac{y-y_F}{z_D-z_F} \tag{17.2}$$

Simple calculations performed with Eqs. 17, 16, and 15 yield the following differential equation which has to be satisfy by the multiplicative inverse of function $f(x,y)$ which generates the transformation of the initial space

$$h_x^2+h_y^2=\frac{1}{z_D^2}\cdot\frac{(x-x_F)^2+(y-y_F)^2}{\varphi^2+(x-x_F)^2+(y-y_F)^2} \tag{18}$$

where $z_D$ and $\varphi=z_D-z_F$ are the thickness and respectively the focal length of the designed lens. The solutions of Eq. 18 will lead us to the transformation function $f(x,y)=1/h(x,y)$ and then to the permittivity and permeability of the designed device given by Eq. 3. Solving Eq. 18 we find the following transformation function

$$h(x,y)=\frac{1}{f(x,y)}=\delta-\gamma\cdot\left[\varphi^2+(x-x_F)^2+(y-y_F)^2\right]^{1/2} \tag{19}$$

where $\gamma=\pm 1/z_D$ and $\delta$ is an integration constant. The anisotropic medium equivalent with the coordinate transformation governed by function $f$, will behave as a converging or diverging lens, depending on a positive or negative value of $\gamma$. If we choose for the integration constant the value $\delta=1+\varphi\cdot\gamma$ then $f(x_F,y_F)=1$, which means that the points located on the lens axis, i.e. the line passing through the focal point $F(x_F, y_F, z_F)$ parallel to $z$-axis, remain unchanged by the coordinate transformation. While the transformation function of the diverging lens ($\gamma<0$) has no singularity points and takes only positive values, the transformation function of the converging lens ($\gamma>0$) has a set of singularity points located on the circle described by the equation $(x-x_F)^2+(y-y_F)^2=z_D\cdot(z_D+2\varphi)$. Inside this circle the function takes positive values, leading to positive values for the constitutive parameters of the equivalent medium, while outside this circle the function takes negative values leading to negative values for the constitutive parameters of the equivalent medium.

The phase accumulated during the propagation of light inside this inhomogeneous and anisotropic media representing the designed flat lens can be easily computed using the following formula

$$\Delta\Phi = \frac{2\pi}{\lambda} \int_p \boldsymbol{k} \cdot d\boldsymbol{r} \qquad (20)$$

where $p$ denotes the wave path inside the lens, $\boldsymbol{k}$ is the dimensionless wave vector and $\lambda$ is the wavelength in free space. Taking into account that in the context of this approach the wave path is a straight line parallel to $z$ axis and the $k_z$ component of the wave vector is constant along the wave path ( $k_z = h$ ) the phase accumulated during propagation inside the designed converging flat lens is given by the following equation

$$\Delta\Phi(x,y) = \frac{2\pi}{\lambda}\left[z_D + \varphi - (\varphi^2 + (x - x_F)^2 + (y - y_F)^2)^{1/2}\right] \qquad (21)$$

Equation 21 shows that the phase shift introduced by the designed converging lens is positive in the space region with positive constitutive parameters, equal to zero in the singularity points, and negative in the space region with negative constitutive parameters of the medium. Also, the phase shift has the same hyperboloidal profile as the phase shift imposed in other approaches of flat lens design [20, 21]. For the particular case of a two-dimensional converging lens the relation between the phase accumulated by the wave propagating inside the lens along its path denoted by $p$ and the coordinate transformation is discussed in section II.D and visually depicted in Fig. 3.

### C. Reflection and transmission

One of the interesting properties of the transformation media is that they are inherently reflectionless. However, this property no longer holds in the case of the so called embedded transformation media, when a finite area of a transformation medium (in this case the lens) is embedded in a different electromagnetic medium (in this case the free space). It was found by heuristic means that the embedded transformation media can be reflectionless if the normal and parallel metrics of the transformed and embedding spaces are identical at interfaces [22]. This topological condition is satisfied at the input interface of the designed device since all the points located on the input plane of the device ($P_0$ plane, see Fig. 1) are not affected by the considered coordinate transformation (see Eq. 1). Hence, at the input interface of the designed lens the incident wave will be totally transmitted inside the device without reflection. The above mentioned topological condition is obviously no longer satisfied at the output interface of the device. Hence, at the output interface of the designed lens the incident wave will split into reflected and transmitted waves.

In this section the reflection and transmission phenomena occurring at the output interface of the lens will be analyzed. In order to compute the reflection ($r$) and transmission ($t$) coefficients it is necessary to describe the incident, reflected and transmitted waves in terms of their wave vector and polarization. While the description of the transmitted wave propagating in the free space is trivial, the description of the incident and reflected waves propagating inside the lens is more complex. The following relations, derived from Maxwell's equations, point out the link between the wave vector and the polarization of a plane wave propagating in a medium having the permittivity $\varepsilon$ and the permeability $\mu$

$$\mu^{-1} \cdot (\boldsymbol{k} \times \boldsymbol{E}) = \boldsymbol{H} \qquad (22.1)$$

$$\boldsymbol{k} \times (\mu^{-1} \cdot (\boldsymbol{k} \times \boldsymbol{E})) + \varepsilon \cdot \boldsymbol{E} = 0 \qquad (22.2)$$

where $\boldsymbol{k}$ is the dimensionless wave vector, $\boldsymbol{E}$ is the magnitude of the electric intensity, and $\boldsymbol{H}$ is the magnitude of the magnetic intensity multiplied by the impedance of free space $\eta_0 = \sqrt{\mu_0/\varepsilon_0}$ [17]. Note that $\boldsymbol{E}$ and $\boldsymbol{H}$ have the same units. Considering that the plane wave propagates in a transformation media ($\varepsilon = \mu = n$) and defining the operator

$$K = \begin{pmatrix} 0 & -k_z & k_y \\ k_z & 0 & -k_x \\ -k_y & k_x & 0 \end{pmatrix} \qquad (23)$$

where $k_x$, $k_y$, and $k_z$ are the components of the $\boldsymbol{k}$ vector on $x$, $y$, and $z$ axis, respectively [19], Eqs. 22 can be expressed as single operators on $\boldsymbol{E}$

$$(n^{-1} \cdot K) \cdot \boldsymbol{E} = P \cdot \boldsymbol{E} = \boldsymbol{H} \qquad (24.1)$$

$$(K \cdot n^{-1} \cdot K + n) \cdot \boldsymbol{E} = M \cdot \boldsymbol{E} = 0 \qquad (24.2)$$

Equations 24 show that the following operators, denoted by $P$ and $M$,

$$P = n^{-1} \cdot K \qquad (25.1)$$

$$M = K \cdot n^{-1} \cdot K + n \qquad (25.2)$$

will define for a given medium, described by $n$, and a plane wave, described by $K$, the polarization of the wave, described by $\boldsymbol{E}$ and $\boldsymbol{H}$.

A normally incident wave on the input plane ($P_0$) of the lens will be totally transmitted inside the medium of the lens without changing its polarization. In the previous sections was proved that this wave will propagate on a linear path parallel to z-axis and the evolution of its wave vector with respect to $z$ coordinate is given by

$$\boldsymbol{k} = (zh_x, zh_y, h)^T \qquad (26)$$

where the superscript $T$ denotes the transposing operation. Plugging the components of vector $\boldsymbol{k}$ (Eq. 26) into the definition of operator $K$ (Eq. 23) and using the relation for tensor $n$ (Eq. 3) the following matrices of operators $P$ and $M$ (Eqs. 25) are obtained

$$P_i = \begin{pmatrix} 0 & -1 & zh_y/h \\ 1 & 0 & -zh_x/h \\ 0 & 0 & 0 \end{pmatrix} \qquad (27.1)$$

$$M_i = \begin{pmatrix} 0 & 0 & 0 \\ 0 & 0 & 0 \\ 0 & 0 & 1/h \end{pmatrix} \qquad (27.2)$$

where the subscript $i$ denotes that these are the operators $P$ and $M$ describing the wave propagating toward the output plane of the lens ($P_1$) which eventually will become the *incident* wave on the output interface of the lens where $z = z_D$. The operators $P_i$ and $M_i$ (Eqs. 27) in conjunction with Eqs. 24 show that the incident wave has no component of the electric or magnetic intensity on z-axis and the *xy* plane components of the electric and magnetic intensities are mutually perpendicular.

In order to study the reflected wave we need first to find its wave vector at the point of reflection on the output plane of the lens ($P_1$). Taking into account that the tangential to the interface components of the wave vectors conserve through reflection and transmission [18], we can easily find that at the reflection point the *x*- and *y*-axis components of the wave vector of the reflected wave are $k_x = z_D h_x$ and $k_y = z_D h_y$ (see Eqs. 15). By plugging these values into the Hamiltonian function given by Eq. 6 and replacing *z* with $z_D$, we obtain the following Hamiltonian function for the output interface of the lens

$$H = (k_z - h)(k_z \cdot (g^2(x,y,z_D)+1) - h \cdot (g^2(x,y,z_D)-1)) \tag{28}$$

where function *g* given by Eq. 4 is evaluated at the output interface of the lens ($z = z_D$). The *z*-axis component of the wave vectors at the output plane of the device can be obtain by requiring the fulfillment of the plane wave dispersion relation represented by the Hamiltonian function given by Eq. 28 $H(k_z) = 0$ [19]. This equation has two solutions: one $k_z = h$ corresponding to the incident wave and another $k_z = h \cdot (g^2(x,y,z_D)-1)/(g^2(x,y,z_D)+1)$ corresponding to the reflected wave. Hence, the wave vector of the reflected wave at the output interface of the lens is

$$\boldsymbol{k}_r = \left(z_D h_x, z_D h_y, h \frac{g_D^2 - 1}{g_D^2 + 1}\right)^T \tag{29}$$

where $g_D = g(x,y,z_D)$ is the *g* function computed at $z = z_D$ and the superscript *T* denotes the transposing operation. Plugging the components of vector $\boldsymbol{k}_r$ (Eq. 29) into the definition of operator *K* (Eq. 23) and using the relation for tensor *n* (Eq. 3) at the output plane of the lens ($z = z_D$) the following matrices of operators *P* and *M* (Eqs. 25) are obtained

$$P_r = \begin{vmatrix} \dfrac{-2 z_D^2 h_x h_y}{g_D^2 + 1} & \dfrac{z_D^2(h_x^2 - h_y^2)+1}{g_D^2 + 1} & \dfrac{z_D h_y}{h} \\ \dfrac{z_D^2(h_x^2 - h_y^2)-1}{g_D^2 + 1} & \dfrac{2 z_D^2 h_x h_y}{g_D^2 + 1} & \dfrac{-z_D h_x}{h} \\ \dfrac{-2 z_D h h_y}{g_D^2 + 1} & \dfrac{2 z_D h h_x}{g_D^2 + 1} & 0 \end{vmatrix} \tag{30.1}$$

$$M_r = \begin{vmatrix} \dfrac{4 z_D^2 h h_x^2}{(g_D^2+1)^2} & \dfrac{4 z_D^2 h h_x h_y}{(g_D^2+1)^2} & \dfrac{-2 z_D h_x}{g_D^2+1} \\ \dfrac{4 z_D^2 h h_x h_y}{(g_D^2+1)^2} & \dfrac{4 z_D^2 h h_y^2}{(g_D^2+1)^2} & \dfrac{-2 z_D h_y}{g_D^2+1} \\ \dfrac{-2 z_D h_x}{g_D^2+1} & \dfrac{-2 z_D h_y}{g_D^2+1} & \dfrac{1}{h} \end{vmatrix} \tag{30.2}$$

where the subscript *r* denotes that these are the operators *P* and *M* describing the reflected wave at the output plane of the lens ($P_1$) where $z = z_D$. The operators $P_r$ and $M_r$ (Eqs. 30) in conjunction with Eqs. 24 show that the reflected wave has non-zero components of electric and magnetic intensities on *z*-axis and the *xy* plane components of the electric and magnetic intensities are mutually perpendicular.

In order to simplify the study of reflection and transmission at the output interface of the lens we will analyze two complementary cases which completely describe the general case. These complementary cases, depicted in figure 2 for a converging lens, are arising from the axial symmetry of the transformed media comprising the lens and are related to the polarization of the incident wave.

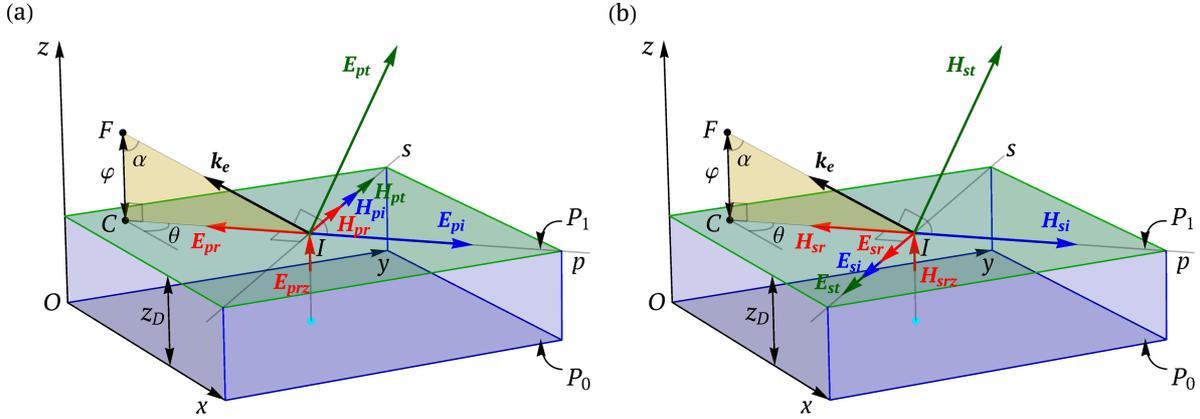

**Figure 2.** The electric and magnetic field intensities of the incident (blue), reflected (red) and transmitted (green) waves at the output interface of the designed converging flat lens. **(a)** *p*-mode, and **(b)** *s*-mode, having the polarization of the incident wave parallel to line *p* and *s*, respectively.

Let *I*, *F*, and *C* denote the point of incidence, the focal point of the lens, and the projection of the focal point on the output interface of the lens, respectively. The lines *IC* and the perpendicular on *IC* passing through *I* are denoted by *p* and *s*, respectively. The angle between the lines *IF* and *CF* is denoted by $\alpha$ and the angle between the line *p* and the *x*-axis is denoted by $\theta$. Figures 2(a) and 2(b) depict the *p*- and *s*-mode, having the polarization of the incident wave along the line *p* and *s*, respectively. The electric and magnetic intensities of the incident, reflected, and transmitted waves are depicted by blue, red, and green colors, respectively. The first subscript denotes the mode (*p* or *s*) and the second subscript denotes the nature of the wave (incident, reflected or transmitted). Before making an in depth analysis of the reflected waves for the *p*- and *s*-modes, we should note that since

$$\sin(\theta) = \frac{y - y_F}{\left((x - x_F)^2 + (y - y_F)^2\right)^{1/2}} \tag{31.1}$$

$$\cos(\theta) = \frac{x - x_F}{\left((x - x_F)^2 + (y - y_F)^2\right)^{1/2}} \tag{31.2}$$

$$\sin(\alpha) = \frac{\left((x - x_F)^2 + (y - y_F)^2\right)^{1/2}}{\left(\varphi^2 + (x - x_F)^2 + (y - y_F)^2\right)^{1/2}} \tag{31.3}$$

the following identities hold for each point *I* from the output plane of the lens ($P_1$) having the coordinates ($x$, $y$, $z_D$)

$$h_x \cos(\theta) + h_y \sin(\theta) = -\gamma g_D \tag{32.1}$$

$$h_x \sin(\theta) = h_y \cos(\theta) \tag{32.2}$$

$$\sin(\alpha)=g_D \tag{32.3}$$

where $h_x$ and $h_y$ are the partial derivatives of function $h$ given by Eq. 19 with respect to $x$ and $y$, respectively and $g_D$ is the function defined by Eq. 4 having $z$ variable set to $z_D$. Using the identities given by Eqs. 32 and the operators $P_r$ and $M_r$ given by Eqs. 30 the components of the electric and magnetic intensities of the reflected wave can be easily computed. For a converging lens when $\gamma=1/z_D$, the electric and magnetic intensities of the $p$-mode are given by

$$\boldsymbol{E}_{pr}=\begin{pmatrix} -E_{pr}\cos(\theta) \\ -E_{pr}\sin(\theta) \\ E_{pr}h\dfrac{2g_D}{g_D^2+1} \end{pmatrix}, \qquad \boldsymbol{H}_{pr}=\begin{pmatrix} -H_{pr}\sin(\theta) \\ H_{pr}\cos(\theta) \\ 0 \end{pmatrix} \tag{33}$$

where $E_{pr}$ and $H_{pr}$ are the $p$-mode magnitudes of the $xy$ plane components of the electric and magnetic intensities of the reflected wave, respectively. Similar relations are obtained for the electric and magnetic intensities of the $s$-mode

$$\boldsymbol{E}_{sr}=\begin{pmatrix} E_{sr}\sin(\theta) \\ -E_{sr}\cos(\theta) \\ 0 \end{pmatrix}, \qquad \boldsymbol{H}_{sr}=\begin{pmatrix} -H_{sr}\cos(\theta) \\ -H_{sr}\sin(\theta) \\ H_{sr}h\dfrac{2g_D}{g_D^2+1} \end{pmatrix} \tag{34}$$

where $E_{sr}$ and $H_{sr}$ are the $s$-mode magnitudes of the $xy$ plane components of the electric and magnetic intensities of the reflected wave, respectively. Relations 33 and 34 show that $xy$ plane components of the electric and magnetic fields of the reflected wave are mutually perpendicular for both $p$- and $s$-mode. Having the complete picture of the reflected wave we can represent the electric and magnetic intensities of the incident, reflected and transmitted waves for $p$- and $s$-mode (see Fig. 2). Note that for the $p$-mode the electric field intensities are in the same plane and the magnetic intensities are collinear, while for the $s$-mode the electric field intensities are collinear and the magnetic intensities are in the same plane.

Considering the conservation of the electric and magnetic intensities parallel to the output interface of the lens [18] the following equations are obtained for $p$-mode (see Fig. 2(a))

$$E_{pi}-E_{pr}=E_{pt}\cos(\alpha) \tag{35.1}$$

$$H_{pi}+H_{pr}=H_{pt} \tag{35.2}$$

and $s$-mode

$$E_{si}+E_{sr}=E_{st} \tag{36.1}$$

$$H_{si}-H_{sr}=H_{st}\cos(\alpha) \tag{36.2}$$

where for $p$- and $s$-mode $\alpha$ is the angle made by the electric and magnetic intensity of the transmitted wave with the line $p$, respectively. Equations similar to 35.2 and 36.1 are obtained when considering the conservation of the electric displacement ($\boldsymbol{D}=\varepsilon\boldsymbol{E}$) and magnetic inductance ($\boldsymbol{B}=\mu\boldsymbol{H}$) perpendicular to the output interface of the lens for $p$- and $s$-mode, respectively. Since in the context of this proof the electric and magnetic intensities have the same units (see Eqs. 22) the following equalities hold $E_{pi}=H_{pi}$, $E_{pr}=H_{pr}$, and $E_{pt}=H_{pt}$ for *p*-mode and $E_{si}=H_{si}$, $E_{sr}=H_{sr}$, and $E_{st}=H_{st}$ for *s*-mode. Taking into account these equalities, Eqs. 35 and 36 yield the same reflection and transmission coefficients for *p*- and *s*-mode given by

$$r = \frac{E_{pr}}{E_{pi}} = \frac{E_{sr}}{E_{si}} = \frac{1-\cos(\alpha)}{1+\cos(\alpha)} \tag{37.1}$$

$$t = \frac{E_{pt}}{E_{pi}} = \frac{E_{st}}{E_{si}} = \frac{2}{1+\cos(\alpha)} \tag{37.2}$$

Similar formulas of reflection and transmission coefficients are obtained for diverging lenses.

### D. Simulation results

The electromagnetic behavior of a convergent lens modeled by the transformation function described by Eq. 19 has been investigated using numerical simulations. For simplicity, the response of the designed flat lens was analyzed for a normally incident transverse magnetic (TM) polarized plane wave having the field components $\{H_x, E_y, E_z\}$. In this particular two-dimensional case the dependence on *x* coordinate can be eliminated from the definition of function *f*. Furthermore, if the focal point is located on *z*-axis ( $x_F = y_F = 0$ ) then the transformation function becomes

$$f(y) = \frac{1}{\delta - \gamma(\varphi^2 + y^2)^{1/2}} \tag{38}$$

where $\gamma = 1/z_D$ ( $z_D$ is the thickness of the lens), $\delta = 1 + \varphi \gamma$ , and $\varphi$ is the focal length of the lens (i.e. the distance from the output plane of the lens to the focal point). Figure 3a depicts the graphs of function *f* and its multiplicative inverse *h*, for a focal length equal with the thickness of the lens ( $\varphi = z_D$ ). This particular transformation function has two singularity points at $\pm y_S$ , where $y_S = z_D \sqrt{3}$ . The function is positive inside the interval $(-y_S, +y_S)$ and negative outside. Since the negative values of the transformation function will produce negative permittivity and permeability, for the current simulations we limited our interest to the positive branch of the transformation function. Figure 3b shows the lines of constant *z* of the original space. The lines corresponding to negative and positive values of *z* are depicted in red and blue, respectively. The area of the original space transformed into the rectangular area of the lens is depicted on figure 3b enclosed by green lines. The left border of this original lens area is given by the equation $z = z_D h(y)$ . Figure 3c shows the lines of constant *z* of the transformed space. After applying the coordinate transformation the original straight lines of constant z are bended according to the shape of the transformation function (see Fig. 3a). The green rectangle depicted on Fig. 3c represents the location and dimension of the lens used in simulations. The lens span on *y*-axis does not exceeds the positive branch area of the transformation function, while on *z*-axis the lens spreads from origin to $z_D$ . An arbitrary wave path inside the lens and its corresponding path in the original space, both denoted by *p*, are depicted on figures 3c and 3b as straight lines parallel to *z*-axis delimited by the borders of the lens area and its corresponding area in the original space, respectively. Using Eq. 20 it is easy to prove that the phase accumulated by the wave propagating inside the lens along the path *p* (see Fig. 3c) is equal to the phase accumulated by a plane wave propagating parallel to *z*-axis along the corresponding original path *p* located in the free space (see Fig. 3b).

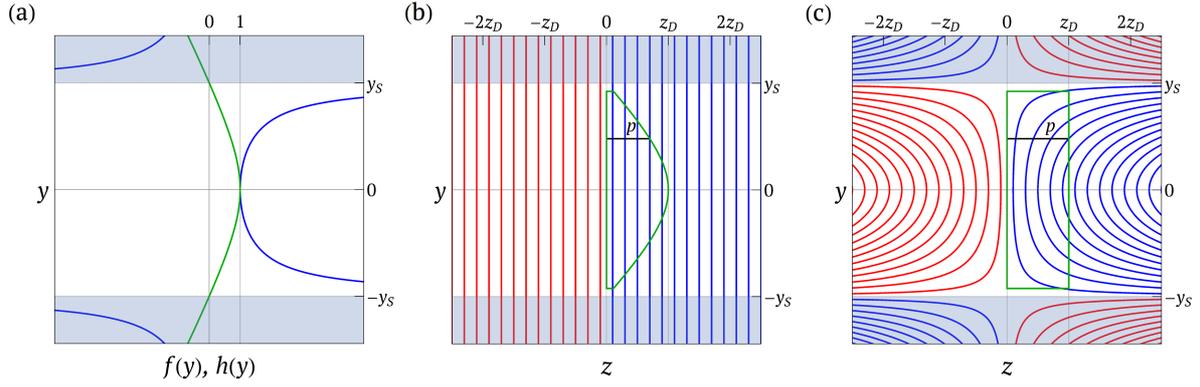

**Figure 3. (a)** The graphs of the transformation function *f* (blue), and its multiplicative inverse *h* (green), used to design a converging flat lens having the thickness equal with the focal length ( $z_D = \varphi$ ). **(b)** The lines of constant positive/negative *z* coordinate of the original space are depicted by blue/red lines parallel to *y*-axis. The original area transformed into the rectangular lens is enclosed by green lines. The original wave path *p*. **(c)** The lines of constant *z* coordinate of the transformed space (the color of the original lines is preserved). The green rectangle shows the location and dimensions of the designed lens. The wave path *p* inside the lens.

The permittivity and permeability tensors describing the medium of the lens are easily obtained by plugging the transformation function given by Eq. 38 into the general form of the constitutive parameters tensor as given by Eq. 3

$$\varepsilon = \mu = \begin{pmatrix} h(y) & 0 & 0 \\ 0 & h(y) & -z h'(y) \\ 0 & -z h'(y) & \dfrac{(z h'(y))^2 + 1}{h(y)} \end{pmatrix} \quad (39)$$

where $h(y) = 1/f(y) = \delta - \gamma \cdot (\varphi^2 + y^2)^{1/2}$ and $h'(y)$ denotes the derivative of function *h*. The numerical simulation have been performed using a two-dimensional Finite-Difference Time-Domain (FDTD) algorithm. In order to avoid violating causality and to assure the convergence of the FDTD algorithm the components of the diagonalized tensor of permittivity and permeability having values less than unity were mapped with a loss free Drude dispersive material model [23]. The input plane of the lens used in simulations was placed in the origin of *z*-axis, and its geometrical dimensions were related to the wavelength of the incident wave as follows: lens thickness $z_D = 5\lambda$ and lens aperture $2 y_D = 16\lambda$ ( $y_D = (8/5) z_D$ satisfies the inequality $y_D < y_S$ ). For these dimensions of the lens, the principal components of the constitutive tensor of the medium comprising the lens are limited by the following values: $0.11 < n_{yy} \leqslant 1$ , $-0.85 < n_{yz} < 0.85$ , and $1 \leqslant n_{zz} < 15.19$ given by Eq. 39, where *n* denotes both permittivity and permeability.

The designed lens was illuminated at normal incidence by a plane wave source of wavelength $\lambda$ with field components $\{E_y, H_x\}$ of magnitude equal to unity located at $z_S = -z_D$ . The rectangular simulation area laid on *y*-axis from $-y_D$ to $y_D$ and on z-axis from $z_S$ to $z_P$ ( $z_P = 5 z_D$ ). The simulation area was terminated with periodic boundary conditions on y-axis and uniaxial perfectly matched layer (UPML) absorbing boundaries on z-axis [24]. The simulations have been performed using an orthogonal computational grid with a step resolution $\Delta = \lambda/50$ for each axes and a time step reaching the Courant limit $\Delta t = \Delta/(c\sqrt{2})$ , where *c* is the speed of light in free space. Computer simulation results presented in figure 4 show the focusing capabilities of the designed flat lens. The real part of the $E_y$ field component along z-axis, on yz plane, and along a line parallel to y-axis passing through the focusing point is depicted in figures 4a, 4b, and 4c, respectively. Figure 4b shows that after passing through the lens the incident plane wave propagates converging to a focal point and after reaching the focal point the wave continues to propagate diverging. The positions of the maximum of the graphs depicted in figures 4a and 4c on z- respectively y-axis show that the focus point resulted from the numerical simulation corresponds in a good approximation with the focal point of the designed lens.

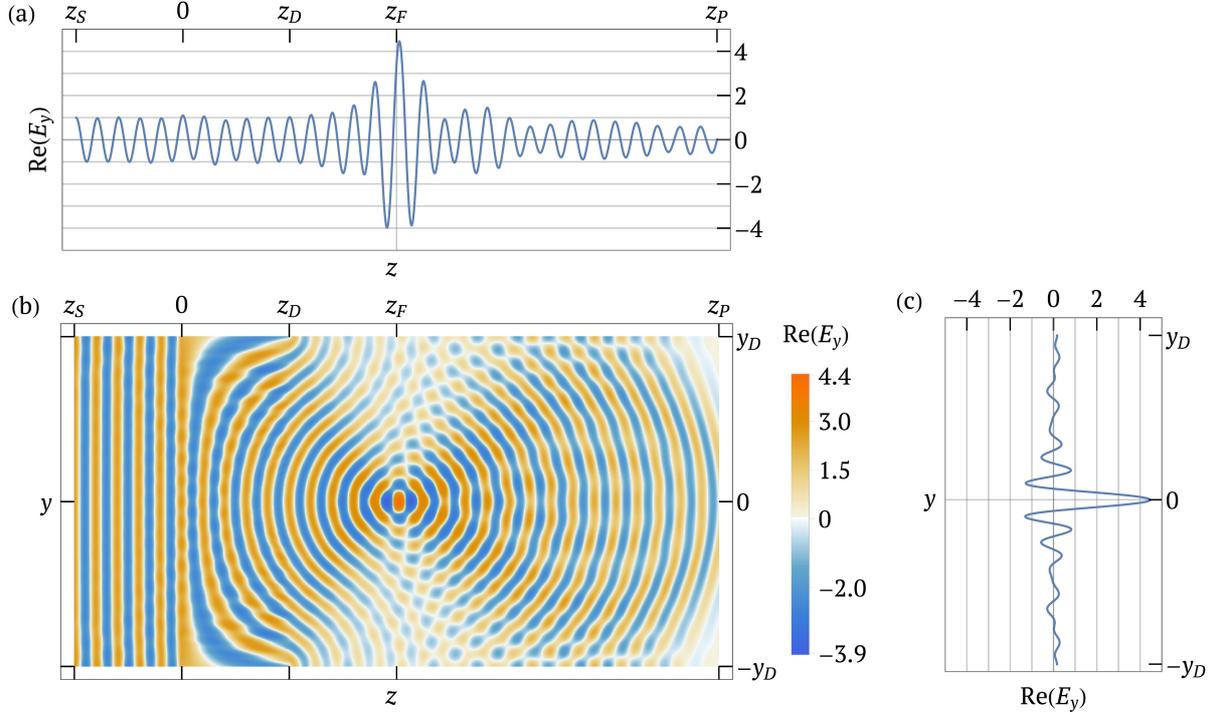

**Figure 4.** The simulated distribution of the real part of the $E_y$ field component. **(a)** Along z-axis, **(b)** on yz plane, and **(c)** along a line parallel to y-axis passing through the focus point resulted from simulation, i.e. the point of the simulation mesh where the real part of $E_y$ reaches its maximum.

In order confirm the theoretical results regarding the reflection and transmission, the electric field intensity resulted from simulation was examined at the output interface of the lens. For the performed two-dimensional simulation the output interface of the lens reduces to the line $z=z_D$ parallel to y-axis. For each point $I$ of the interface, having the coordinates $(0, y, z_D)$ the angle $\alpha$ satisfies the following equation $\sin(\alpha)=|y|/(\varphi^2+y^2)^{1/2}$. Since the polarization of the incident wave is parallel to y-axis, the performed simulation corresponds to a p-mode. The magnitude of the $E_y$ field component *inside* the lens next to the interface represents the value of $E_{pi}-E_{pr}$, while the magnitudes of the $E_y$ and $E_z$ field components *outside* the lens next to the interface will give the electric intensity of the transmitted wave $E_{pt}=(|E_y|^2+|E_z|^2)^{1/2}$. Using Eq. 37.1 the following relation is obtained for the overlapping fields of the incident and reflected waves

$$1-r=\frac{E_{pi}-E_{pr}}{E_{pi}}=\frac{2\cos(\alpha)}{1+\cos(\alpha)} \qquad (40)$$

Since the magnitude of the electric intensity of the incident wave used in simulation is equal to unity ($E_{pi}=1$), the values of $E_{pi}-E_{pr}$ and $E_{pt}$, resulted from simulation, should be compared with the values of $1-r$ and $t$ given by Eq. 40 and Eq. 37.2, respectively. The blue graphs depicted on figures 5(a) and (b) represent the values of $E_{pi}-E_{pr}$ and $E_{pt}$ resulted from simulation, respectively. The brown graphs depicted on figures 5(a) and (b) represent the functions $1-r$ and $t$ computed for the output interface of the lens, respectively.

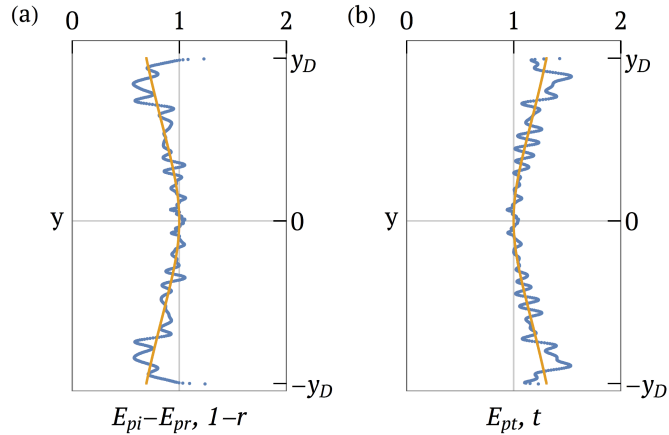

**Figure 5. (a)** The simulated values of $E_{pi}-E_{pr}$ inside the lens at the output interface (blue) and the computed values of $1-r$ at the output interface of the lens (brown). **(b)** The simulated values of $E_{pt}$ *outside* the lens at the output interface (blue) and the computed values of $t$ at the output interface of the lens (brown).

The graphs depicted in figure 5 show that the simulated values of the electric field inside and outside the lens at its output interface fit well with the theoretically deducted formulas of reflection and transmission.

### III. Conclusion

In this study we have demonstrated that the concept of transformation optics can be involved in designing electromagnetic devices capable primarily to manipulate the wave vector of the probing electromagnetic wave in the desired manner, while the wave path remains of a secondary importance. Considering a specific class of coordinate transformation which alters only one coordinate with respect to the other two coordinates of the original system we have proved that devices having a trivial geometry, a medium separated by two parallel planes, can be designed in order to fulfill different electromagnetic functions sustained by wave vector alteration. The desired modification of the wave vector component perpendicular to the device axis will be obtained at normal incidence by choosing the appropriate transformation function which satisfies a specific differential equations.

These theoretical results have been applied for the design of converging and diverging flat lenses. Computer simulations show reliable focusing capabilities of the converging flat lens designed in the context of the developed theory. The study of reflection and transmission shows that the designed converging lens may focus very well in the mid-near field and far field regions corresponding to sufficiently small *α* angles. When the focal point is placed far from the output plane of the lens the area of the transformation medium with positive permittivity and permeability expands so that the implementation of the designed lens needs only materials with positive optical properties.

Other devices able to focus or defocus the electromagnetic radiation in or from a specific area of the space can be easily conceived and designed due to the flexibility and simplicity of wave vector manipulation technique provided by these theoretical approach. Such devices may fulfill useful functions in imaging systems, information processing systems or cloaking devices.

**Acknowledgments:**

This work was developed by the Company for Applied Informatics using infrastructure obtained through the program POSCCE-A2-O2.3.2-2012-1 (ID: 1542/SMIS: 44023), and partially supported by the program PNII/2014-nr. 153/2014.